\documentclass[12pt]{article}
\usepackage{amsmath,amssymb,amsthm,amsxtra,overpic,bm,epsfig,subfigure}
\usepackage{color}
\usepackage{comment}
\usepackage[all,cmtip]{xy}
\usepackage{multirow}
\usepackage{hyperref}
\usepackage{graphicx}
\usepackage{dcolumn}
\usepackage{url}            
\usepackage{booktabs}       
\usepackage{amsfonts}       
\usepackage{nicefrac}       
\usepackage{microtype}      
\usepackage{lipsum}
\usepackage{mathrsfs}
\usepackage{indentfirst}
\usepackage{float}
\usepackage[utf8]{inputenc}
\usepackage{cite}
\usepackage[table]{xcolor}

\usepackage{slashed,stmaryrd}

\renewcommand{\thefootnote}{\fnsymbol{footnote}}

\begin{document}

\vspace{0.2cm}

\begin{center}
{\large\bf Further study on the lepton mass spectra and flavor mixing with $S_{3L} \times S_{3R}$ flavor symmetry}
\end{center}

\vspace{0.2cm}

\begin{center}
{\bf Chang-Qi Hu}\textsuperscript{a} \footnote{Corresponding author. E-mail:23412011092@stumail.sdut.edu.cn}, 
{\bf Chun-Yuan Li}\textsuperscript{a} \footnote{Corresponding author. E-mail:lichunyuan@sdut.edu.cn},
{\bf Xiao Liang}\textsuperscript{a} \footnote{Corresponding author. E-mail:lx321@sdut.edu.cn},
{\bf Xing-Hua Yang}\textsuperscript{a} \footnote{Corresponding author. E-mail:yangxinghua@sdut.edu.cn}, 
{\bf Dai-Xing Zhang}\textsuperscript{a} \footnote{Corresponding author. E-mail:24412011130@stumail.sdut.edu.cn}
\\
\textsuperscript{a}{School of Physics and Optoelectronic Engineering, Shandong University of Technology, Zibo, Shandong 255000, China}\\

\end{center}

\vspace{1.5cm}

\begin{abstract}

Neutrino oscillation experiments have confirmed that neutrinos are massive particles and lepton flavors are mixed. To explain the observed lepton mass spectra and flavor mixing patterns, flavor symmetry plays a crucial and unique role.
In this paper, we propose a useful symmetry-breaking scheme by applying $S_{3L} \times S_{3R} \rightarrow S_{2L} \times S_{2R} \rightarrow \emptyset$ within both charged-lepton and neutrino sectors at the mass-matrix level.
For the three distinct residual subgroups $S_{2L}^{(23)} \times S_{2R}^{(23)}$, $S_{2L}^{(13)} \times S_{2R}^{(13)}$ and $S_{2L}^{(12)} \times S_{2R}^{(12)}$ under consideration, we systematically analyze the various parameterizations of the lepton mass matrices. It is shown that all the three scenarios are in good agreement with current neutrino oscillation data. Notably, within the latest best-fit values of neutrino oscillation parameters, the predicted Dirac CP-violating phase $\delta$ is calculated to be $294.6^\circ$, $302.3^\circ$ and $287.0^\circ$, respectively.
To further assess the viability of the model, a comprehensive numerical analysis is performed by utilizing neutrino oscillation parameters at the $3\sigma$ level. It is found that the allowed range of $\delta$ is $281.2^\circ \rightarrow 338.7^\circ$, $287.0^\circ \rightarrow 342.2^\circ$ and $282.7^\circ \rightarrow 297.0^\circ$, all fall within its $3\sigma$ range.
These results indicate that the proposed symmetry-breaking scheme $S_{3L} \times S_{3R} \rightarrow S_{2L} \times S_{2R} \rightarrow \emptyset$ can naturally explain the realistic lepton mass hierarchy and mixing pattern, thereby providing valuable theoretical perspectives for future research.

\end{abstract}

\def\thefootnote{\arabic{footnote}}
\setcounter{footnote}{0}

\newpage

\section{Introduction}\label{sec1}

The standard model is one of the most successful theories in physics, providing a precise and unified description of fundamental particles and their interactions~\cite{Glashow:1961tr,Weinberg:1967tq,Salam:1968rm}. However, there are still several unresolved problems~\cite{DiBari:2021fhs,Arbey:2021gdg,Crivellin:2023zui,Losada:2021bxx}. Among these open questions, the phenomena of neutrino oscillations provide the most direct evidence of new physics beyond the standard model~\cite{SajjadAthar:2021prg}. Within the framework of the standard model, the absence of right-handed neutrino fields prevents the generation of Dirac mass terms through Yukawa couplings, and thus neutrinos remain strictly massless. In recent years, a series of neutrino oscillation experiments have revealed that neutrinos are massive particles and lepton flavors are mixed, which has stimulated extensive research on new physics beyond the standard model, see Ref.~\cite{Denton:2025jkt} for recent review.

Recently, the neutrino oscillation experiments, such as SNO+~\cite{SNO:2025koj}, Super-Kamiokande~\cite{Super-Kamiokande:2023jbt}, T2K~\cite{T2K:2023mcm} and NO$\nu$A~\cite{NOvA:2025tmb} have significantly improved the precision of neutrino oscillation parameter measurements, including three mixing angles ($\theta_{12}, \theta_{23}, \theta_{13}$) and two mass-squared differences ($\Delta m^2_{21},\Delta m^2_{31}$).
The global analysis of neutrino oscillation data also provides constraints on the observable parameters~\cite{Esteban:2024eli,Capozzi:2025wyn,deSalas:2020pgw}.
For instance, the latest best-fit values of three neutrino mixing angles provided by NuFIT 6.0 (2024)~\cite{Esteban:2024eli} are
\begin{align}\label{1}
 \theta_{12} = 33.68^{ +0.73^{\circ}}_{-0.70^{\circ}} \; , \quad
 \theta_{23} = 48.5^{ +0.7^{\circ}}_{-0.9^{\circ}}\; , \quad
 \theta_{13} = 8.52^{ + 0.11^{\circ}}_{-0.11^\circ}\; .
\end{align}
In case of the normal neutrino mass hierarchy $(m_1 < m_2 < m_3)$, the best-fit values of two neutrino mass-squared differences are
\begin{align}\label{2}
\Delta m^2_{21} = (7.49^{+0.19}_{-0.19}) \times 10^{-5}~\text{eV}^2 \; , \quad
\Delta m^2_{31} = (2.534^{+0.025}_{-0.023}) \times 10^{-3}~\text{eV}^2  \; .
\end{align}
While the neutrino mass-squared differences are measured with ever-increasing precision, the absolute scale of neutrino masses has not yet been determined.
The KATRIN experiment recently performs precision spectroscopy of the tritium $\beta$-decay close to the kinematic endpoint and sets an upper limit on the effective electron antineutrino mass of $m_\nu < 0.45 \, \text{eV}$ at $90\%$ confidence level~\cite{KATRIN:2024cdt}.

The above discussions strongly indicate that the generation mechanism of neutrino masses is quite different from that of other fermions in the standard model. At present, various mechanisms for neutrino mass generation have been proposed, such as the canonical seesaw mechanism~\cite{Minkowski:1977sc,Yanagida:1979as,Gell-Mann:1979vob,Glashow:1979nm,Mohapatra:1979ia}, inverse seesaw mechanism~\cite{Wyler:1982dd,Mohapatra:1986bd}, and radiative mass generation~\cite{Branco:1978bz,Chang:1986bp,Babu:1988yq,Hung:1998tv}. While effective in generating small neutrino masses, these models generally lack the ability to constrain the flavor structures of massive neutrinos, which usually requires additional flavor symmetries~\cite{Xing:2020ijf, King:2013eh}.
In the literature, the $S_{3L} \times S_{3R}$ symmetry has been widely studied, as it predicts that the Yukawa interactions of charged-fermions take the well-known ``democratic'' matrix form~\cite{Harari:1978yi,Froggatt:1978nt,Koide:1989ds,Tanimoto:1989qh,Kaus:1990ij,Branco:1990fj,Fritzsch:1989qm,Fritzsch:1994yx,Branco:1995pw,Fritzsch:1995dj,Xing:1996hi,Mondragon:1998gy,Fritzsch:1998xs,Haba:2000rf,Branco:2001hn,Fujii:2002jw,Fritzsch:2004xc,Rodejohann:2004qh,Teshima:2005bk,Altarelli:2010gt,Ishimori:2010au,Xing:2010iu,Zhou:2011nu,Dev:2012ns,GonzalezCanales:2012blg,Jora:2012nw,Ishimori:2013woa,Yang:2016esx,Fritzsch:2017tyf,Si:2017pdo,Pramanick:2019oxb,Mishra:2019sye,Mishra:2019keq,Garcia-Aguilar:2021xgk,Babu:2023oih,Li:2024pff,Gomez-Izquierdo:2024apr,Gresnigt:2024awl}.
Furthermore, in order to account for the mass spectra and flavor mixing, a variety of different symmetry-breaking patterns have also been proposed.

In this paper, we adopt a phenomenological approach and work at the mass-matrix level. In order to explain the realistic lepton mass spectra and mixing angles, we apply the permutation symmetry $S_3$ to both charged-lepton and neutrino  sectors. Different from previous studies, we propose an interesting scheme to break the $S_{3L} \times S_{3R}$ flavor symmetry, where the same two-stage symmetry-breaking chain $S_{3L} \times S_{3R} \rightarrow S_{2L} \times S_{2R} \rightarrow \emptyset$ is applied in parallel to charged-lepton and neutrino mass matrices. There are a total of nine parameters in the model, which can be fully determined by nine corresponding observables. Moreover, the model enables the predictions of the Dirac and Majorana CP-violating phases, the sum of neutrino masses, the effective neutrino mass as well as other relevant quantities.

To be explicit, the remaining part of this paper is organized as follows. A theoretical framework based on $S_{3L} \times S_{3R}$ flavor symmetry is briefly introduced in Sec.~\ref{sec:method}. In Sec.~\ref{sec:analysis}, we specify the three distinct symmetry-breaking chains and explore their implications for the lepton mass spectra and flavor mixing. Sec.~\ref{sec:numerics} details the numerical determination of model parameters. Finally, we present the brief summary in Sec.~\ref{sec:conclusion}.

\section{$S_{3L} \times S_{3R}$ Flavor Symmetry Framework}\label{sec:method}

As indicated in Ref.\cite{Zhou:2011nu}, from a phenomenological perspective, the Lagrangian relevant to lepton masses at low energies can be expressed as
\begin{equation}\label{3}
-\mathcal{L}_{\text{mass}}= \overline{l_L} M_l l_R + \frac{1}{2} \overline{\nu_L} M_\nu (\nu^c_L) + \text{h.c.} \; ,
\end{equation}
where $M_l$ denotes the mass matrix of the charged-leptons, and $M_\nu$ represents the effective Majorana neutrino mass matrix. The latter may naturally arise from various neutrino mass models.
To account for the experimentally observed lepton mass spectra and flavor mixing patterns, we reconsider a simple model based on the discrete symmetry $S_{3L} \times S_{3R}$.
Here, the subscripts $L$ and $R$ denote the transformation properties under the left- and right-handed flavor symmetry, respectively.
$S_3$ denotes the permutation group of three objects, and its three-dimensional reducible representation can be expressed as
\begin{align}\label{4}
S^{(123)} &=
\begin{pmatrix}
1 & 0 & 0 \\
0 & 1 & 0 \\
0 & 0 & 1
\end{pmatrix}, &
S^{(231)} &=
\begin{pmatrix}
0 & 1 & 0 \\
0 & 0 & 1 \\
1 & 0 & 0
\end{pmatrix},\notag \\
S^{(312)} &=
\begin{pmatrix}
0 & 0 & 1 \\
1 & 0 & 0 \\
0 & 1 & 0
\end{pmatrix}, &
S^{(213)} &=
\begin{pmatrix}
0 & 1 & 0 \\
1 & 0 & 0 \\
0 & 0 & 1
\end{pmatrix},\notag  \\
S^{(132)} &=
\begin{pmatrix}
1 & 0 & 0 \\
0 & 0 & 1 \\
0 & 1 & 0
\end{pmatrix}, &
S^{(321)} &=
\begin{pmatrix}
0 & 0 & 1 \\
0 & 1 & 0 \\
1 & 0 & 0
\end{pmatrix}.
\end{align}

Assuming the Lagrangian in Eq.~\ref{3} remains invariant under $S_{3L} \times S_{3R}$ symmetry, then the mass matrices of the charged-leptons and neutrinos should satisfy the following relations
\begin{align}\label{5}
S_{3L}M_l = M_l S_{3R}  \; ,  \quad
S_{3L}M_\nu = M_\nu S_{3L} \; .
\end{align}
Specifically, in the limit of $S_{3L} \times S_{3R}$ symmetry, the charged-lepton mass matrix takes the so-called democratic form
\begin{equation}\label{6}
M^{(0)}_l= \frac{c_l}{3}
\begin{pmatrix}
1 & 1 & 1 \\
1 & 1 & 1 \\
1 & 1 & 1
\end{pmatrix},
\end{equation}
where $c_l$ measures the mass scale of charged-leptons.
Similarly, the neutrino mass matrix in the symmetry limit can be expressed as
\begin{equation}\label{7}
M^{(0)}_\nu = c_\nu\left[\begin{pmatrix}
1 & 0 & 0 \\
0 & 1 & 0 \\
0 & 0 & 1
\end{pmatrix} + r_\nu \begin{pmatrix}
1 & 1 & 1 \\
1 & 1 & 1 \\
1 & 1 & 1
\end{pmatrix}\right],
\end{equation}
where $c_\nu$ sets the mass scale of neutrinos, and $r_\nu$ quantifies the deviation from the identity matrix.
It is worth mentioning that in order to get two large lepton mixing angles, $|r_\nu| \ll 1$ must be satisfied.
Note that both $M^{(0)}_l$ and $M^{(0)}_\nu$ can be diagonalized by the same orthogonal matrix as follows
\begin{equation}\label{8}
V_D^T M_l^{(0)} V_D = \frac{c_l}{3}
\begin{pmatrix}
0 & 0 & 0 \\
0 & 0 & 0 \\
0 & 0 & 3
\end{pmatrix},\;
V_D^T M_\nu^{(0)} V_D = c_\nu
\begin{pmatrix}
1 & 0 & 0 \\
0 & 1 & 0 \\
0 & 0 & 1 + 3r_\nu
\end{pmatrix},\;
\text{with} \;
V_D =
\begin{pmatrix}
\frac{1}{\sqrt{2}} & \frac{1}{\sqrt{6}} & \frac{1}{\sqrt{3}} \\
-\frac{1}{\sqrt{2}} & \frac{1}{\sqrt{6}} & \frac{1}{\sqrt{3}} \\
0 & -\frac{2}{\sqrt{6}} & \frac{1}{\sqrt{3}}
\end{pmatrix}.
\end{equation}
It is obvious from Eq.~\ref{8} that only the third-generation charged-lepton acquires a non-zero mass, and the first two generations remain massless. In the neutrino sector, the three generation neutrinos are nearly degenerate. Furthermore, the lepton mixing matrix in this case turns out to be an identity matrix.
Hence, it is imperative to break the $S_{3L} \times S_{3R}$ symmetry in the charged-lepton and neutrino mass matrices to explain the realistic lepton masses and mixing angles.

In this work, we assume that the $S_{3L} \times S_{3R}$ symmetry may first be broken to a residual $S_{2L} \times S_{2R}$ symmetry, and followed by random perturbations to entirely break the remaining symmetry. The two-stage symmetry-breaking scheme is applied to both charged-lepton and neutrino mass matrices in a similar way.
According to the above discussions, the mass matrices of the charged-leptons and neutrinos can then be decomposed as
\begin{align}\label{9}
M_l = M^{(0)}_l + \Delta M^{(1)}_l + \Delta M^{(2)}_l \; ,
\end{align}
\begin{align}\label{10}
M_\nu = M^{(0)}_\nu + \Delta M^{(1)}_\nu + \Delta M^{(2)}_\nu \; .
\end{align}
Here $M^{(0)}_l$ and $M^{(0)}_\nu$ are the symmetry-limit terms determined by the $S_{3L} \times S_{3R}$ symmetry, as shown in Eq.~\ref{6} and Eq.~\ref{7}.
$\Delta M^{(1)}_l$ and $\Delta M^{(1)}_\nu$ are the first-order perturbation terms controlled by the residual $S_{2L} \times S_{2R}$ symmetry, while $\Delta M^{(2)}_l$ and $\Delta M^{(2)}_\nu$ are the second-order perturbation terms that employed to eventually break the remaining symmetry.
The specific form of the perturbation terms will be shown later.

\section{Lepton Mass Spectra and Mixing Matrix}\label{sec:analysis}

Based on the symmetry-breaking hypothesis proposed above, the first-order perturbation terms $\Delta M^{(1)}_l$ and $\Delta M^{(1)}_\nu$ should be $S_{2L} \times S_{2R}$ invariant in our model.
According to the different group elements of the $S_{2L} \times S_{2R}$ symmetry, three different breaking patterns are investigated in detail, which can be classified as
\begin{align}\label{11}
{\rm Scenario~I:} \quad
S_{3L} \times S_{3R} &\rightarrow S^{(23)}_{2L} \times S^{(23)}_{2R} \rightarrow \emptyset \; ,
\text {with} ~~ S^{(23)}_2 = \{ S^{(123)},~S^{(132)} \} \; ,
\end{align}
\begin{align}\label{12}
{\rm Scenario~II:} \quad
S_{3L} \times S_{3R} &\rightarrow S^{(13)}_{2L} \times S^{(13)}_{2R} \rightarrow \emptyset\; ,
\text {with} ~~ S^{(13)}_2 = \{ S^{(123)},~S^{(321)} \} \; ,
\end{align}
\begin{align}\label{13}
{\rm Scenario~III:} \quad
S_{3L} \times S_{3R} &\rightarrow S^{(12)}_{2L} \times S^{(12)}_{2R} \rightarrow \emptyset\; ,
\text {with} ~~ S^{(12)}_2 = \{ S^{(123)},~S^{(213)} \} \; .
\end{align}

As there is no residual symmetry left in the charged-lepton and neutrino sector, the second-order perturbation terms $\Delta M^{(2)}_l$ and $\Delta M^{(2)}_\nu$ are usually random anarchy. For simplicity, $\Delta M^{(2)}_l$ and $\Delta M^{(2)}_\nu$ are chosen to be diagonal in our later discussions.

\subsection{Scenario I ($S_{3L} \times S_{3R} \rightarrow S^{(23)}_{2L} \times S^{(23)}_{2R} \rightarrow \emptyset$)}

For the symmetry-breaking chain $S_{3L} \times S_{3R} \rightarrow S^{(23)}_{2L} \times S^{(23)}_{2R} \rightarrow \emptyset$, the general form of the first-order perturbation term $\Delta M^{(1)}_l$ can be given by

\begin{equation}\label{14}
\Delta M^{(1)}_l = \frac{c_l}{3} \begin{pmatrix}
\delta_l & 0 & 0 \\
0 & \delta_l & \delta_l \\
0 & \delta_l & \delta_l
\end{pmatrix},
\end{equation}
where \(\delta_l\) is a small real parameter and \( |\delta_l| \ll 1 \).
Since the residual $S_{2L} \times S_{2R}$ symmetry is eventually broken, there is no remaining symmetry in the charged-lepton mass matrix, and the perturbations in second-order term $\Delta M^{(2)}_l$ are typically random.
However, at least from the point of view of model building, it is more natural to consider simple forms.
For simplicity, the following diagonal form are selected
\begin{equation}\label{15}
\Delta M^{(2)}_l = \frac{c_l}{3}
\begin{pmatrix}
-i\epsilon_l & 0 & 0 \\
0 & i\epsilon_l & 0 \\
0 & 0 & \varepsilon_l
\end{pmatrix}.
\end{equation}
Here \( \varepsilon_l \), \( \epsilon_l \) are both small perturbative parameters.
As the symmetry-breaking terms in charged-lepton sector are responsible for the generation of muon and electron masses, $|\delta_l|, |\epsilon_l| \ll |\varepsilon_l| < 1$ are required.
It is worthwhile to mention that here $\Delta M^{(2)}_l$ is assumed to be a complex matrix, which is intentional since we expect appropriate CP violation in the lepton sector.

Now, the mass matrix of charged-lepton in Eq.~\ref{9} can be written as
\begin{align}\label{16}
M_l &=\frac{c_l}{3}\left[
\begin{pmatrix}
1 & 1 & 1 \\
1 & 1 & 1 \\
1 & 1 & 1
\end{pmatrix}+
\begin{pmatrix}
\delta_l & 0 & 0 \\
0 & \delta_l & \delta_l \\
0 & \delta_l & \delta_l
\end{pmatrix}+
\begin{pmatrix}
-i\epsilon_l & 0 & 0 \\
0 & i\epsilon_l & 0 \\
0 & 0 & \varepsilon_l
\end{pmatrix}\right] \notag \\
&= \frac{c_l}{3} \begin{pmatrix}
1 + \delta_l - i \epsilon_l & 1 & 1 \\
1 & 1+\delta_l + i \epsilon_l & 1+\delta_l \\
1 & 1+\delta_l & 1+\delta_l+\varepsilon_l
\end{pmatrix},
\end{align}
in which $c_l >0$ and $|\delta_l|, |\epsilon_l| \ll |\varepsilon_l| < 1$. Note that here $M_l$ is a complex symmetric matrix, and can usually be diagonalized by a unitary matrix.
After diagonalizing $M_l$ through $V_l^\dagger M_l V_l^\ast = \text{Diag}\{m_e, m_\mu, m_\tau\}$, the masses of three charged-leptons can be obtained as
\begin{equation}\label{17}
m_\tau \approx c_l \left(1 + \frac{\varepsilon_l}{9} + \frac{5\delta_l}{9} \right) , \
m_\mu \approx c_l \left(  \frac{2\varepsilon_l}{9} + \frac{\delta_l}{9} \right) , \
m_e \approx c_l \left( \frac{\delta_l}{3} + \frac{\epsilon_l^2}{6\varepsilon_l} \right) \; .
\end{equation}
Furthermore, the unitary matrix $V_l$ is found to be
\begin{align}\label{18}
V_l \approx & \frac{1}{\sqrt{6}}
\begin{pmatrix}
\sqrt{3} & 1 & \sqrt{2} \\
-\sqrt{3} & 1 &  \sqrt{2} \\
0 & -2 & \sqrt{2}
\end{pmatrix}
\!+\!  \frac{i\epsilon_l}{2\sqrt{2}\varepsilon_l}
\begin{pmatrix}
-1 & -\sqrt{3} & 0 \\
-1 & \sqrt{3} & 0 \\
2 & 0 & 0
\end{pmatrix}
 \!+\! \frac{\varepsilon_l}{9\sqrt{3}}
\begin{pmatrix}
0 & \sqrt{2} & -1 \\
0 & \sqrt{2} & -1 \\
0 & \sqrt{2} & 2  \\
\end{pmatrix}.
\end{align}
It is worth mentioning that the first term in $V_l$ is just the orthogonal matrix $V_D$ that used to diagonalize the $S_{3L} \times S_{3R}$ symmetry-limit terms $M^{(0)}_l$ and $M^{(0)}_\nu$, as shown in Eq.~\ref{8}.

In the neutrino sector, the first-order perturbation term $\Delta M^{(1)}_\nu$ with residual $S^{(23)}_{2L} \times S^{(23)}_{2R}$ symmetry can be expressed as
\begin{equation}\label{19}
\Delta M^{(1)}_\nu = c_\nu
\begin{pmatrix}
\delta_\nu & 0 & 0 \\
0 & 0 & \delta_\nu \\
0 & \delta_\nu & 0
\end{pmatrix},
\end{equation}
where \( |\delta_\nu| \ll 1 \).
Similar to the charged-lepton sector, there is no residual symmetry left in the neutrino sector, which allows the second-order perturbation term $\Delta M^{(2)}_\nu$ to be arbitrary. 
It is worth mentioning that as the neutrinos are assumed to be Majorana particles, the neutrino mass matrix should be symmetric.
For simplicity, $\Delta M^{(2)}_\nu$ is assumed to take the following diagonal form
\begin{equation}\label{20}
\Delta M^{(2)}_\nu = c_\nu
\begin{pmatrix}
-\epsilon_\nu & 0 & 0 \\
0 & \epsilon_\nu & 0 \\
0 & 0 & \varepsilon_\nu
\end{pmatrix}.
\end{equation}
Since the symmetry-breaking terms in neutrino sector are responsible for the breaking of neutrino mass degeneracy, the perturbation parameters should satisfy $|\delta_\nu|, |\epsilon_\nu| \ll |\varepsilon_\nu| < 1$.
The general form of neutrino mass matrix $M_\nu$ with perturbations can then be read as
\begin{align}\label{21}
M_\nu &= c_\nu\left[\begin{pmatrix}
1 & 0 & 0 \\
0 & 1 & 0 \\
0 & 0 & 1
\end{pmatrix} + r_\nu \begin{pmatrix}
1 & 1 & 1 \\
1 & 1 & 1 \\
1 & 1 & 1
\end{pmatrix} +
\begin{pmatrix}
\delta_\nu & 0 & 0 \\
0 & 0 & \delta_\nu \\
0 & \delta_\nu & 0
\end{pmatrix} +
\begin{pmatrix}
-\epsilon_\nu & 0 & 0 \\
0 & \epsilon_\nu & 0 \\
0 & 0 & \varepsilon_\nu
\end{pmatrix}\right] \notag \\
&= c_\nu
\begin{pmatrix}
1 + r_\nu + \delta_\nu - \epsilon_\nu & r_\nu & r_\nu \\
r_\nu & 1 + r_\nu + \epsilon_\nu & r_\nu + \delta_\nu \\
r_\nu & r_\nu + \delta_\nu & 1 + r_\nu + \varepsilon_\nu
\end{pmatrix},
\end{align}
where $c_\nu >0$ and $|r_\nu|, |\delta_\nu|, |\epsilon_\nu| \ll |\varepsilon_\nu| < 1$ are implied. Note that the neutrino mass matrix $M_\nu$ derived in this case is the same as that given in Ref.~\cite{Zhou:2011nu}.
After diagonalizing $M_\nu$ via $V_\nu^\dagger M_\nu V_\nu^\ast = \text{Diag}\{m_1, m_2, m_3\}$, the three neutrino mass eigenvalues can be given by
\begin{equation}\label{22}
\begin{split}
m_3 &\approx c_\nu (1 + r_\nu + \varepsilon_\nu ) \; , \\
m_2 &\approx c_\nu \left( 1 + r_\nu + \frac{1}{2}\delta_\nu + \frac{1}{2} \sqrt{(2\epsilon_\nu-\delta_\nu)^2+4r_\nu^2} \right) \; , \\
m_1 &\approx c_\nu \left( 1 + r_\nu + \frac{1}{2}\delta_\nu - \frac{1}{2} \sqrt{(2\epsilon_\nu-\delta_\nu)^2+4r_\nu^2} \right) \; .
\end{split}
\end{equation}
It is easy to find that the neutrino mass spectrum in Eq.~\ref{22} shows a normal mass hierarchy $(m_1 < m_2 < m_3)$.
Approximately, the unitary matrix \( V_\nu \) can be expressed as
\begin{equation}\label{23}
V_\nu \approx \frac{1}{\varepsilon_\nu}
\begin{pmatrix}
\varepsilon_\nu c_\theta & \varepsilon_\nu s_\theta & r_\nu \\
-\varepsilon_\nu s_\theta & \varepsilon_\nu c_\theta & r_\nu + \delta_\nu \\
(r_\nu + \delta_\nu) s_\theta - r_\nu c_\theta & -(r_\nu + \delta_\nu) c_\theta - r_\nu s_\theta & \varepsilon_\nu
\end{pmatrix},
\end{equation}
where $c_\theta \equiv \cos\theta$, $s_\theta \equiv \sin\theta$ with $\tan2\theta = 2r_\nu/(2\epsilon_\nu - \delta_\nu)$.

The lepton mixing matrix arises from the mismatch between the diagonalization of the charged-lepton mass matrix $M_l$ and that of the neutrino mass matrix $M_\nu$, and can be defined as $V_{\text{PMNS}}=V_l^\dagger V_\nu$.
According to Eq.~\ref{18} and Eq.~\ref{23}, the lepton mixing matrix $V_{\text{PMNS}}$ can then be approximately expressed as
\begin{align}\label{24}
V_{\rm PMNS}&=\frac{1}{\sqrt{6}}\left(\begin{array}{ccc}
\sqrt{3}(c_\theta+s_\theta) & \sqrt{3}(s_\theta-c_\theta) & 0 \\
c_\theta-s_\theta & c_\theta+s_\theta & -2 \\
\sqrt{2}(c_\theta-s_\theta) & \sqrt{2}(c_\theta+s_\theta) & \sqrt{2} \\
\end{array}\right) \notag \\
&+\frac{i\epsilon_l}{2\sqrt{2}\varepsilon_l} \left(\begin{array}{ccc}
c_\theta-s_\theta & (c_\theta+s_\theta) & -2\\
\sqrt{3}(c_\theta+s_\theta) & \sqrt{3}(s_\theta-c_\theta) & 0 \\
0 & 0 & 0 \\
\end{array}\right) \notag \\
&+\frac{\varepsilon_l}{9\sqrt{3}} \left(\begin{array}{ccc}
0 & 0 & 0 \\
\sqrt{2}(c_\theta-s_\theta) & \sqrt{2}(c_\theta+s_\theta) & \sqrt{2} \\
s_\theta-c_\theta & -(c_\theta+s_\theta) & 2\\
\end{array}\right) \notag \\
&+\frac{r_\nu}{\sqrt{6}\varepsilon_\nu} \left(\begin{array}{ccc}
0 & 0 & 0 \\
2(c_\theta-s_\theta) & 2(c_\theta+s_\theta) & 2\\
\sqrt{2}(s_\theta-c_\theta) & -\sqrt{2}(c_\theta+s_\theta) & 2\sqrt{2} \\
\end{array}\right) \notag \\
&+\frac{\delta_\nu}{\sqrt{6}\varepsilon_\nu}\left(\begin{array}{ccc}
0 & 0 & -\sqrt{3} \\
-2s_\theta & 2c_\theta & 1\\
\sqrt{2}s_\theta & -\sqrt{2}c_\theta & \sqrt{2} \\
\end{array}\right)\ .
\end{align}

\subsection{Scenario II ($S_{3L} \times S_{3R} \rightarrow S^{(13)}_{2L} \times S^{(13)}_{2R} \rightarrow \emptyset$)}

In case of the symmetry-breaking chain is taken as $S_{3L} \times S_{3R} \rightarrow S^{(13)}_{2L} \times S^{(13)}_{2R} \rightarrow \emptyset$,
the first-order perturbation term $\Delta M^{(1)}_l$ can be obtained as
\begin{equation}\label{25}
\Delta M^{(1)}_l=\frac{c_l}{3}\left(
\begin{array}{ccc}
\delta_l & 0 & \delta_l \\
0 & \delta_l & 0 \\
\delta_l & 0 & \delta_l \\
\end{array}
\right),
\end{equation}
with $|\delta_l| \ll 1$.
On the other hand, the second-order perturbation term $\Delta M^{(2)}_l$ is also assumed to be diagonal and takes the same form as in Eq.~\ref{15}.
Finally, the charged-lepton mass matrix $M_l$ can be explicitly expressed as
\begin{align}\label{26}
M_l &=\frac{c_l}{3}\left[
\begin{pmatrix}
1 & 1 & 1 \\
1 & 1 & 1 \\
1 & 1 & 1
\end{pmatrix}+
\begin{pmatrix}
\delta_l & 0 & \delta_l \\
0 & \delta_l & 0 \\
\delta_l & 0 & \delta_l
\end{pmatrix}+
\begin{pmatrix}
-i\epsilon_l & 0 & 0 \\
0 & i\epsilon_l & 0 \\
0 & 0 & \varepsilon_l
\end{pmatrix}\right] \notag \\
&= \frac{c_l}{3} \begin{pmatrix}
1+\delta_l - i\epsilon_l & 1 & 1+\delta_l \\
1 & 1 + \delta_l + i\epsilon_l & 1 \\
1+\delta_l & 1 & 1+\delta_l + \varepsilon_l
\end{pmatrix},
\end{align}
where $c_l >0$ and $|\delta_l|, |\epsilon_l| \ll |\varepsilon_l| < 1$ are assumed.
The mass matrix $M_l$ of charged-lepton obtained in Eq.~\ref{26} is also complex symmetric, and can be diagonalized by a unitary matrix.
Through a simple perturbation calculation, the masses of three charged-leptons are approximately given by
\begin{equation}\label{27}
m_\tau \approx c_l \left(1 + \frac{\varepsilon_l}{9} + \frac{5\delta_l}{9}\right) , \
m_\mu \approx c_l \left(  \frac{2\varepsilon_l}{9} + \frac{\delta_l}{9} \right) , \
m_e \approx c_l \left( \frac{\delta_l}{3} + \frac{\epsilon_l^2}{6\varepsilon_l} \right) \; .
\end{equation}
Note that the charged-lepton mass spectrum obtained here is exactly the same as that given in Eq.~\ref{17}.
The corresponding unitary matrix \(V_l\) can be expressed as
\begin{align}\label{28}
    V_l \approx & \frac{1}{\sqrt{6}}
\begin{pmatrix}
\sqrt{3} & 1 & \sqrt{2} \\
-\sqrt{3} & 1 &  \sqrt{2} \\
0 & -2 & \sqrt{2}
\end{pmatrix}
\!+\!  \frac{i\epsilon_l}{2\sqrt{2}\varepsilon_l}
\begin{pmatrix}
-1 & -\sqrt{3} & 0 \\
-1 & \sqrt{3} & 0 \\
2 & 0 & 0
\end{pmatrix} \!+\! \frac{\varepsilon_l}{9\sqrt{3}}
\begin{pmatrix}
0 & \sqrt{2} & -1 \\
0 & \sqrt{2} & -1 \\
0 & \sqrt{2} & 2  \\
\end{pmatrix}.
\end{align}

For the neutrino sector, the first-order perturbation term $\Delta M^{(1)}_\nu$ in the limit of the residual $S^{(13)}_{2L} \times S^{(13)}_{2R}$ symmetry can be given by
\begin{equation}\label{29}
\Delta M^{(1)}_\nu = c_\nu
\begin{pmatrix}
0 &  0 &\delta_\nu \\
0 & \delta_\nu & 0 \\
\delta_\nu & 0 & 0
\end{pmatrix},
\end{equation}
with \( |\delta_\nu| \ll 1 \).
As given in Eq.~\ref{20}, the second-order perturbation term $\Delta M^{(2)}_\nu$ is also chosen to be diagonal for simplicity.
The mass matrix $M_\nu$ of neutrino can now be written as
\begin{align}\label{30}
M_\nu &= c_\nu\left[\begin{pmatrix}
1 & 0 & 0 \\
0 & 1 & 0 \\
0 & 0 & 1
\end{pmatrix} + r_\nu \begin{pmatrix}
1 & 1 & 1 \\
1 & 1 & 1 \\
1 & 1 & 1
\end{pmatrix} +
\begin{pmatrix}
0 &  0 &\delta_\nu \\
0 & \delta_\nu & 0 \\
\delta_\nu & 0 & 0
\end{pmatrix} +
\begin{pmatrix}
-\epsilon_\nu & 0 & 0 \\
0 & \epsilon_\nu & 0 \\
0 & 0 & \varepsilon_\nu
\end{pmatrix}\right] \notag \\
&= c_\nu
\begin{pmatrix}
1 + r_\nu - \epsilon_\nu & r_\nu & r_\nu + \delta_\nu \\
r_\nu & 1 + r_\nu + \delta_\nu + \epsilon_\nu & r_\nu  \\
r_\nu + \delta_\nu & r_\nu & 1 + r_\nu + \varepsilon_\nu
\end{pmatrix},
\end{align}
where $c_\nu >0$ and $|r_\nu|, |\delta_\nu|, |\epsilon_\nu| \ll |\varepsilon_\nu| < 1$.
By diagonalizing the above matrix, we obtain the mass eigenvalues of three neutrinos with high accuracy
\begin{equation}\label{31}
\begin{split}
m_3&\approx c_\nu(1+r_\nu+\varepsilon_\nu)\ , \\
m_2&\approx c_\nu\left(1+r_\nu+\frac{1}{2}\delta_\nu+ \frac{1}{2} \sqrt{(2\epsilon_\nu+\delta_\nu)^2+4r_\nu^2}\right)\ , \\
m_1&\approx c_\nu\left(1+r_\nu+\frac{1}{2}\delta_\nu- \frac{1}{2} \sqrt{(2\epsilon_\nu+\delta_\nu)^2+4r_\nu^2}\right)\ .
\end{split}
\end{equation}
The unitary matrix used to diagonalize $M_\nu$ via $V_\nu^\dagger M_\nu V_\nu^\ast = \text{Diag}\{m_1, m_2, m_3\}$ is found
to be
\begin{equation}\label{32}
V_\nu\approx \frac{1}{\varepsilon_\nu}\left(
\begin{array}{ccc}
\varepsilon_\nu c_\theta & \varepsilon_\nu s_\theta & r_\nu+\delta_\nu \\
-\varepsilon_\nu s_\theta & \varepsilon_\nu c_\theta & r_\nu \\
r_\nu s_\theta-(r_\nu+\delta_\nu) c_\theta &  -(r_\nu+\delta_\nu) s_\theta-r_\nu  c_\theta & \varepsilon_\nu \\
\end{array}
\right),
\end{equation}
with \(\tan2\theta=2r_\nu/(2\epsilon_\nu+\delta_\nu)\).

In this case, the explicit form of the lepton mixing matrix $V_{\rm PMNS} = V^\dagger_l V_\nu$ can be approximately derived as
\begin{align}\label{33}
V_{\rm PMNS}&=\frac{1}{\sqrt{6}}\left(\begin{array}{ccc}
\sqrt{3}(c_\theta+s_\theta) & \sqrt{3}(s_\theta-c_\theta) & 0 \\
c_\theta-s_\theta & c_\theta+s_\theta & -2 \\
\sqrt{2}(c_\theta-s_\theta) & \sqrt{2}(c_\theta+s_\theta) & \sqrt{2} \\
\end{array}\right) \notag \\
&+\frac{i\epsilon_l}{2\sqrt{2}\varepsilon_l} \left(\begin{array}{ccc}
c_\theta-s_\theta & (c_\theta+s_\theta) & -2\\
\sqrt{3}(c_\theta+s_\theta) & \sqrt{3}(s_\theta-c_\theta) & 0 \\
0 & 0 & 0 \\
\end{array}\right) \notag \\
&+\frac{\varepsilon_l}{9\sqrt{3}} \left(\begin{array}{ccc}
0 & 0 & 0 \\
\sqrt{2}(c_\theta-s_\theta) & \sqrt{2}(c_\theta+s_\theta) & \sqrt{2} \\
s_\theta-c_\theta & -(c_\theta+s_\theta) & 2\\
\end{array}\right) \notag \\
&+\frac{r_\nu}{\sqrt{6}\varepsilon_\nu} \left(\begin{array}{ccc}
0 & 0 & 0 \\
2(c_\theta-s_\theta) & 2(c_\theta+s_\theta) & 2\\
\sqrt{2}(s_\theta-c_\theta) & -\sqrt{2}(c_\theta+s_\theta) & 2\sqrt{2} \\
\end{array}\right) \notag \\
&+\frac{\delta_\nu}{\sqrt{6}\varepsilon_\nu}\left(\begin{array}{ccc}
0 & 0 & \sqrt{3} \\
2c_\theta & 2s_\theta & 1\\
-\sqrt{2}c_\theta & -\sqrt{2}s_\theta & \sqrt{2} \\
\end{array}\right)\ .
\end{align}

\subsection{Scenario III ($S_{3L} \times S_{3R} \rightarrow S^{(12)}_{2L} \times S^{(12)}_{2R} \rightarrow \emptyset$)}

When the symmetry-breaking chain is assumed to be $S_{3L} \times S_{3R} \rightarrow S^{(12)}_{2L} \times S^{(12)}_{2R} \rightarrow \emptyset$, the first-order perturbation term $\Delta M^{(1)}_l$ is found to be
\begin{equation}\label{34}
\Delta M^{(1)}_l = \frac{c_l}{3}
\begin{pmatrix}
\delta_l & \delta_l & 0 \\
\delta_l & \delta_l & 0 \\
0 & 0 & \delta_l
\end{pmatrix},
\end{equation}
with $|\delta_l| \ll 1$.
The second-order perturbation term $\Delta M^{(2)}_l$ is also set as a diagonal matrix, as shown in Eq.~\ref{15}.
According to the above discussions, the mass matrix $M_l$ of the charged-lepton can be expressed as
\begin{align}\label{35}
M_l &=\frac{c_l}{3}\left[
\begin{pmatrix}
1 & 1 & 1 \\
1 & 1 & 1 \\
1 & 1 & 1
\end{pmatrix}+
\begin{pmatrix}
\delta_l & \delta_l & 0 \\
\delta_l & \delta_l & 0 \\
0 & 0 & \delta_l
\end{pmatrix}+
\begin{pmatrix}
-i\epsilon_l & 0 & 0 \\
0 & i\epsilon_l & 0 \\
0 & 0 & \varepsilon_l
\end{pmatrix}\right] \notag \\
&= \frac{c_l}{3} \begin{pmatrix}
1+\delta_l-i \epsilon_l & 1+\delta_l & 1 \\
 1+\delta_l & 1+\delta_l+i\epsilon_l & 1 \\
1 & 1 & 1+ \delta_l +\varepsilon_l
\end{pmatrix},
\end{align}
where $c_l >0$ and $|\delta_l|, |\epsilon_l| \ll |\varepsilon_l| < 1$.
By diagonalizing the charged-lepton mass matrix $M_l$, we get the masses of three charged-leptons
\begin{align}\label{36}
 m_\tau \approx c_l \left(1 + \frac{\varepsilon_l}{9} + \frac{5\delta_l}{9} \right),\
 m_\mu \approx c_l \left(  \frac{2\varepsilon_l}{9} + \frac{4\delta_l}{9} \right) ,\
 m_e \approx c_l \left(  \frac{\epsilon_l^2}{6\varepsilon_l} - \frac{\delta_l\epsilon_l^2}{3\varepsilon^2_l} \right) .
\end{align}
The unitary matrix $V_l$ used to diagonalize $M_l$ is given by
\begin{align}\label{37}
    V_l \approx \frac{1}{\sqrt{6}}
\begin{pmatrix}
\sqrt{3} & 1 & \sqrt{2} \\
-\sqrt{3} & 1 &  \sqrt{2} \\
0 & -2 & \sqrt{2}
\end{pmatrix}
\!+\!  \frac{i\epsilon_l}{2\sqrt{2}\varepsilon_l}
\begin{pmatrix}
-1 & -\sqrt{3} & 0 \\
-1 & \sqrt{3} & 0 \\
2 & 0 & 0
\end{pmatrix}
 \!+\! \frac{\varepsilon_l}{9\sqrt{3}}
\begin{pmatrix}
0 & \sqrt{2} & -1 \\
0 & \sqrt{2} & -1 \\
0 & \sqrt{2} & 2  \\
\end{pmatrix}.
\end{align}

Turning to the neutrino sector, the first-order perturbation term $\Delta M^{(1)}_\nu$ constrained by the residual $S^{(12)}_{2L} \times S^{(12)}_{2R}$ symmetry is found to be
\begin{equation}\label{38}
\Delta M^{(1)}_\nu = c_\nu
\begin{pmatrix}
0 & \delta_\nu & 0 \\
\delta_\nu & 0 & 0 \\
0 & 0 & \delta_\nu
\end{pmatrix},
\end{equation}
with $|\delta_\nu| \ll 1$.
The second-order perturbation term $\Delta M^{(2)}_\nu$ is set to the same form as in Eq.~\ref{20}.
The mass matrix $M_\nu$ of neutrino can then be expressed as
\begin{align}\label{39}
M_\nu &= c_\nu\left[\begin{pmatrix}
1 & 0 & 0 \\
0 & 1 & 0 \\
0 & 0 & 1
\end{pmatrix} + r_\nu \begin{pmatrix}
1 & 1 & 1 \\
1 & 1 & 1 \\
1 & 1 & 1
\end{pmatrix} +
\begin{pmatrix}
0 & \delta_\nu & 0 \\
\delta_\nu & 0 & 0 \\
0 & 0 & \delta_\nu
\end{pmatrix} +
\begin{pmatrix}
-\epsilon_\nu & 0 & 0 \\
0 & \epsilon_\nu & 0 \\
0 & 0 & \varepsilon_\nu
\end{pmatrix}\right] \notag \\
&= c_\nu
\begin{pmatrix}
1 + r_\nu - \epsilon_\nu & r_\nu + \delta_\nu & r_\nu  \\
r_\nu  + \delta_\nu& 1 + r_\nu + \epsilon_\nu & r_\nu  \\
r_\nu & r_\nu  & 1 + r_\nu + \delta_\nu + \varepsilon_\nu
\end{pmatrix},
\end{align}
where $c_\nu >0$ and $|r_\nu|, |\delta_\nu|, |\epsilon_\nu| \ll |\varepsilon_\nu| < 1$.
Diagonalizing the above matrix \( M_\nu \), the three neutrino mass eigenvalues can be derived as
\begin{equation}\label{40}
\begin{split}
m_3 &\approx c_\nu (1 + \delta_\nu + r_\nu + \varepsilon_\nu), \\
m_2 &\approx c_\nu \left(1 + r_\nu + \sqrt{(\delta_\nu + r_\nu)^2 +\epsilon_\nu^2}\right), \\
m_1 &\approx c_\nu \left(1 + r_\nu - \sqrt{(\delta_\nu + r_\nu)^2 +\epsilon_\nu^2}\right).
    \end{split}
\end{equation}
The unitary matrix used to diagonalize the neutrino mass matrix can be approximately given by
\begin{equation}\label{41}
V_\nu \approx \frac{1}{\varepsilon_\nu}
\begin{pmatrix}
\varepsilon_\nu c_\theta & \varepsilon_\nu s_\theta & r_\nu \\
-\varepsilon_\nu s_\theta & \varepsilon_\nu c_\theta & r_\nu  \\
r_\nu (s_\theta - c_\theta) & -r_\nu (c_\theta + s_\theta) & \varepsilon_\nu
\end{pmatrix},
\end{equation}
with \(\tan2\theta=(r_\nu+\delta_\nu)/\epsilon_\nu\).
From Eq.~\ref{37} and Eq.~\ref{41}, we can derive the lepton mixing matrix $V_{\rm PMNS} = V^\dagger_l V_\nu$. More explicitly,
\begin{align}\label{42}
V_{\rm PMNS}&=\frac{1}{\sqrt{6}}\left(\begin{array}{ccc}
\sqrt{3}(c_\theta+s_\theta) & \sqrt{3}(s_\theta-c_\theta) & 0 \\
c_\theta-s_\theta & c_\theta+s_\theta & -2 \\
\sqrt{2}(c_\theta-s_\theta) & \sqrt{2}(c_\theta+s_\theta) & \sqrt{2} \\
\end{array}\right) \notag \\
&+\frac{i\epsilon_l}{2\sqrt{2}\varepsilon_l} \left(\begin{array}{ccc}
c_\theta-s_\theta & (c_\theta+s_\theta) & -2\\
\sqrt{3}(c_\theta+s_\theta) & \sqrt{3}(s_\theta-c_\theta) & 0 \\
0 & 0 & 0 \\
\end{array}\right) \notag \\
&+\frac{\varepsilon_l}{9\sqrt{3}} \left(\begin{array}{ccc}
0 & 0 & 0 \\
\sqrt{2}(c_\theta-s_\theta) & \sqrt{2}(c_\theta+s_\theta) & \sqrt{2} \\
s_\theta-c_\theta & -(c_\theta+s_\theta) & 2\\
\end{array}\right) \notag \\
&+\frac{r_\nu}{\sqrt{6}\varepsilon_\nu} \left(\begin{array}{ccc}
0 & 0 & 0 \\
2(c_\theta-s_\theta) & 2(c_\theta+s_\theta) & 2\\
\sqrt{2}(s_\theta-c_\theta) & -\sqrt{2}(c_\theta+s_\theta) & 2\sqrt{2} \\
\end{array}\right) .
\end{align}

At this stage, we have derived the lepton mixing matrix corresponding to the three distinct symmetry-breaking chains.
In these scenarios, various symmetry-breaking patterns give rise to different parameterizations of the lepton mass matrices, which ultimately leads to distinct lepton mixing matrix.
By comparing the obtained lepton mixing matrix with its standard parametrization, one can easily extract three neutrino mixing angles and the CP-violating phase, which we will elaborate on in next Section.

\section{Numerical Analysis and Predictions for CP Violation}\label{sec:numerics}

In general, the lepton mixing matrix $V_{\rm PMNS}$ can be parameterized by three mixing angles ($\theta_{12}$, $\theta_{13}$, $\theta_{23}$) and one Dirac CP-violating phase $\delta$. If neutrinos are Majorana particles, $V_{\rm PMNS}$ also includes two additional Majorana phases $\rho$ and $\sigma$.
The standard parameterization of $V_{\rm PMNS}$ can be expressed as~\cite{ParticleDataGroup:2024cfk}
\begin{equation}\label{43}
V_{\text{PMNS}} =
\begin{pmatrix}
c_{13}c_{12} & c_{13}s_{12} & s_{13}e^{-i\delta} \\
-s_{12}c_{23}-c_{12}s_{13}s_{23}e^{i\delta} & c_{12}c_{23}-s_{12}s_{13}s_{23}e^{i\delta} & c_{13}s_{23} \\
s_{12}s_{23}-c_{12}s_{13}c_{23}e^{i\delta} & -c_{12}s_{23}-s_{12}s_{13}c_{23}e^{i\delta} & c_{13}c_{23}
\end{pmatrix}\cdot P_{\rm M} ,
\end{equation}
where $c_{ij} \equiv \cos\theta_{ij}$, $s_{ij} \equiv \sin\theta_{ij}$ and $P_{\rm M}={\rm Diag}\{e^{i \rho}, e^{i\sigma}, 1\}$.
Based on the previous discussions, it is evident that there are nine parameters in our model, that is, four parameters ($c_l, \delta_l, \epsilon_l, \varepsilon_l$) in the charged-lepton sector and five parameters ($c_\nu, r_\nu, \delta_\nu, \epsilon_\nu, \varepsilon_\nu$) in the neutrino sector.
These parameters can be fully determined by nine corresponding observables, namely, three charged-lepton masses ($m_e, m_\mu, m_\tau$ ), three neutrino masses ($m_1, m_2,m_3$), and three neutrino mixing angles ($\theta_{12}, \theta_{13}, \theta_{23}$). On this basis, we can further give the predicted value of the CP-violating phase.
For convenience, the latest best-fit values of neutrino oscillation parameters for normal neutrino mass hierarchy provided by NuFIT 6.0 (2024)~\cite{Esteban:2024eli} are adopted in the following numerical analysis.

For Scenario I, the two neutrino mass-squared differences can be calculated from Eq.~\ref{22}
\begin{equation}\label{44}
\Delta m^2_{31} \approx c^2_\nu \varepsilon_\nu(2+\varepsilon_\nu) \; , \quad
\Delta m^2_{21} \approx 2c^2_\nu \sqrt{(2\epsilon_\nu-\delta_\nu)^2+4r_\nu^2} \; .
\end{equation}
The three neutrino masses in Eq.~\ref{22} are nearly degenerate, indicating that the effective neutrino mass in tritium beta decays and that in neutrinoless double-beta decays are of the same order as the neutrino mass scale parameter $c_\nu$.
In the conservative case, $c_\nu \approx 0.03~{\rm eV}$ is taken, which is consistent  with the current cosmological observations~\cite{Planck:2018vyg}.
By using the best-fit value of $\Delta m^2_{31} = 2.534 \times 10^{-3}~\text{eV}^2$, one can get
\begin{equation}\label{45}
\varepsilon_\nu(2+\varepsilon_\nu) \approx \frac{\Delta m^2_{31}}{c^2_\nu} \rightarrow \varepsilon_\nu \approx 0.953 \; .
\end{equation}

By comparing the lepton mixing matrix $V_{\rm PMNS}$ in Eq.~\ref{24} with the standard parametrization in Eq.~\ref{43}, one can find that
\begin{equation}\label{46}
\sin^2{\theta_{12}}=\frac{|V_{e2}|^2}{1-|V_{e3}|^2} \approx \frac{1}{2}(1-\sin{2\theta}) \; .
\end{equation}
Making use of the best-fit value of $\theta_{12} \approx 33.68^\circ$, we get $\theta = 11.32^\circ$.
With the help of Eq.~\ref{44} and the relation $\tan2\theta = 2r_\nu/(2\epsilon_\nu - \delta_\nu)$ defined before, we can obtain
\begin{equation}\label{47}
\frac{2r_\nu}{\varepsilon_\nu(2+\varepsilon_\nu)} = \cos\theta \sin\theta \frac{\Delta m^2_{21}}{\Delta m^2_{31}} \; .
\end{equation}
By inputting the best-fit value of $\Delta m^2_{21} = 7.49 \times 10^{-5}~\text{eV}^2$, one can then derive
\begin{equation}\label{48}
r_\nu \approx 8.01\times 10^{-3}  \; .
\end{equation}

Adopting the standard parametrization in Eq.~\ref{43}, we can also get
\begin{align}\label{49}
\sin^2\theta_{23} & \approx \frac{|V_{\mu3}|^2}{1-|V_{e3}|^2} \approx
\frac{2}{3} \left( 1 - \frac{2\varepsilon_l}{9} - \frac{2r_\nu}{\varepsilon_\nu} -  \frac{\delta_\nu}{\varepsilon_\nu}\right) \; , \notag \\
\sin^2\theta_{13} &\approx|V_{e3}|^2 \approx
\frac{1}{2}\left(\frac{\delta_\nu^2}{\varepsilon_\nu^2}+ \frac{\epsilon_l^2}{\varepsilon_l^2}\right)  \; .
\end{align}
Combining Eq.~\ref{17} and Eq.~\ref{49}, one finally find
\begin{equation}\label{50}
\varepsilon_l \approx 0.275 \; , \quad \epsilon_l \approx -5.31\times 10^{-2} \; , \quad\delta_l \approx -4.27 \times 10^{-3}\; , \quad  \delta_\nu\approx 7.69 \times 10^{-2} \; ,
\end{equation}
by inputting the best-fit value of $\theta_{23} = 48.5^\circ, \theta_{13} = 8.52^\circ$, together with the charged-lepton masses $m_e = 0.488 \text{MeV}, m_\mu = 102.877 \text{MeV}$ and $m_\tau = 1747.43 \text{MeV}$ at the electroweak scale~\cite{Huang:2020hdv}.
The magnitude of $\epsilon_\nu$ can be obtained through
\begin{equation}\label{51}
\epsilon_\nu = \frac{1}{2}\left(\frac{2r_\nu}{\tan2\theta} + \delta_\nu\right) \approx 5.76\times 10^{-2} \; .
\end{equation}
Through the above calculations, it is easy to find that all the model parameters are in good agreement with our initial expectations that $|\delta_l|, |\epsilon_l| \ll |\varepsilon_l| < 1$ and $|r_\nu|, |\delta_\nu|, |\epsilon_\nu| \ll |\varepsilon_\nu| < 1$.

The Dirac CP-violating phase $\delta$ can be extracted from the Jarlskog invariant~\cite{Jarlskog:1985ht,Wu:1985ea}
\begin{equation}\label{52}
{\cal J}= \rm{Im[V_{\mu1}V_{\tau2}V^*_{\mu2}V^*_{\tau1}]}=\frac{1}{8}\cos{\theta_{13}}\sin{2\theta_{12}}\sin{2\theta_{23}}\sin{2\theta_{13}}\sin{\delta} \; .
\end{equation}
It can be found that the predicted value of $\delta$ is approximately
\begin{equation}\label{53}
\delta \approx 294.6^\circ \; ,
\end{equation}
which falls within the $3\sigma$ interval and will be tested in future neutrino oscillation experiments.
There is no doubt that the numerical prediction of the Dirac CP-violating phase presented here depends on the set-up of the second-order perturbation term in previous section.

The two Majorana CP-violating phases $\rho$ and $\sigma$ can also be extracted as
\begin{equation}\label{54}
\rho=\arctan\left[{\frac{1}{2}\frac{\epsilon_l}{\varepsilon_l}\frac{c_\theta-s_\theta}{c_\theta+s_\theta}}\right]\approx -3.68^\circ \; , \quad
\sigma=\arctan\left[{\frac{1}{2}\frac{\epsilon_l}{\varepsilon_l}\frac{c_\theta+s_\theta}{s_\theta-c_\theta}}\right] \approx 8.24^\circ \ ,
\end{equation}
which are close to the trivial value.

Finally, we present the sum of neutrino masses $\sum{m_\nu}$ and the effective neutrino mass $\langle m \rangle_{ee}$ of the neutrinoless double-beta decay. As a result,
\begin{align}
\label{55}
\sum{m_\nu} &= m_1+m_2+m_3\approx0.12~\rm{eV}  \; , \notag \\
\langle m \rangle_{ee} &= \left|\sum_i{m_i V^2_{ei}}\right| \approx 2.91\times10^{-2}~\rm{eV} \; .
\end{align}

Following the same procedure as used for Scenario I, we can also calculate the values of the model parameters in Scenario II and Scenario III, and provide the predicted value of Dirac CP-violating phase.
In Table~\ref{tab}, we list the values of the model parameters and the predicted Dirac CP-violating phase for each scenario.
It turns out that all the three scenarios are consistent well with current experimental
data, and the predicted values of Dirac CP-violating phase all lie within $3\sigma$ range.

\begin{table}[!htbp]
\caption{The values of the model parameters and the predicted Dirac CP-violating phase for each scenario.}\label{tab}
\centering
\resizebox{\linewidth}{!}{
\begin{tabular}{|c|c|c|}
\hline
Scenario & Model parameters  & Dirac CP-violating phase \\
\hline
\multirow{2}{*}{Scenario~I} &\multirow{1}{*}{$\varepsilon_l \approx 0.275,~\delta_l \approx -4.27\times 10^{-3},~\epsilon_l \approx -5.31\times10^{-2}$} & \multirow{2}{*}{$\delta \approx 294.6^\circ$} \\
 & \multirow{1}{*}{$\varepsilon_\nu \approx 0.953,~r_\nu \approx 8.01\times 10^{-3},~\delta_\nu \approx 7.69\times 10^{-2},~\epsilon_\nu \approx 5.76\times 10^{-2}$}  & \\
\hline
\multirow{2}{*}{Scenario~II} & \multirow{1}{*}{$\varepsilon_l \approx 0.275,~\delta_l \approx -4.27\times 10^{-3},~\epsilon_l \approx -5.31\times10^{-2}$} & \multirow{2}{*}{$\delta \approx 302.3^\circ$}\\
 & \multirow{1}{*}{$\varepsilon_\nu \approx 0.953,~r_\nu \approx 8.01\times 10^{-3},~\delta_\nu \approx 7.69\times 10^{-2},~\epsilon_\nu \approx -1.93 \times 10^{-2}$} & \\
\hline
\multirow{2}{*}{Scenario~III} & \multirow{1}{*}{$\varepsilon_l \approx 0.159,~\delta_l \approx 5.96\times10^{-2},~\epsilon_l \approx -3.34\times10^{-2} $} & \multirow{2}{*}{$\delta \approx 287.0^\circ$} \\
 & \multirow{1}{*}{$\varepsilon_\nu \approx 0.953,~r_\nu \approx 5.87\times 10^{-2},~\delta_\nu \approx -6.67 \times 10^{-2} ,~\epsilon_\nu \approx -1.92 \times 10^{-2}$} & \\
\hline
\end{tabular}}
\end{table}

In order to assess the validity of the model more comprehensively, the $3\sigma$ ranges of three neutrino mixing angles and two mass-squared differences from the latest global analysis of neutrino oscillation data are implemented~\cite{Esteban:2024eli}
\begin{align}\label{57}
\theta_{12} = 31.63^{\circ} &\rightarrow 35.95^{\circ} \; , \quad
\theta_{13} = 8.18^{\circ} \rightarrow 8.87^{\circ} \; , \quad
\theta_{23} = 41.0^{\circ} \rightarrow 50.5^{\circ} \; , \notag \\
\Delta m^2_{21} = (6.92 &\rightarrow8.05 )\times 10^{-5}\text{eV}^2 \; , \quad
\Delta m^2_{31} = (2.463 \rightarrow2.606 ) \times 10^{-3}\text{eV}^2 \; .
\end{align}
For the three scenarios, the corresponding predictions of the Dirac CP-violating phase $\delta$ can be respectively given by
\begin{align}\label{58}
\delta &= 281.2^\circ \rightarrow 338.7^\circ \quad \text{for Scenario I} \; , \notag \\
\delta &= 287.0^\circ \rightarrow 342.2^\circ \quad \text{for Scenario II} \; , \notag\\
\delta &= 282.7^\circ \rightarrow 297.0^\circ \quad \text{for Scenario III} \; ,
\end{align}
which all lie within the $3\sigma$ range.
For the sake of example, the variations of the predicted Dirac CP-violating phase $\delta$ with $\sin^2\theta_{23}$ for Scenario I, Scenario II and Scenario III are shown in Figure~\ref{fig}, respectively.

\setcounter{figure}{0} 
\begin{figure}[!htbp]
\centering
\subfigure[Scenario I]{\label{fig1a}
\includegraphics[width=0.3\textwidth]{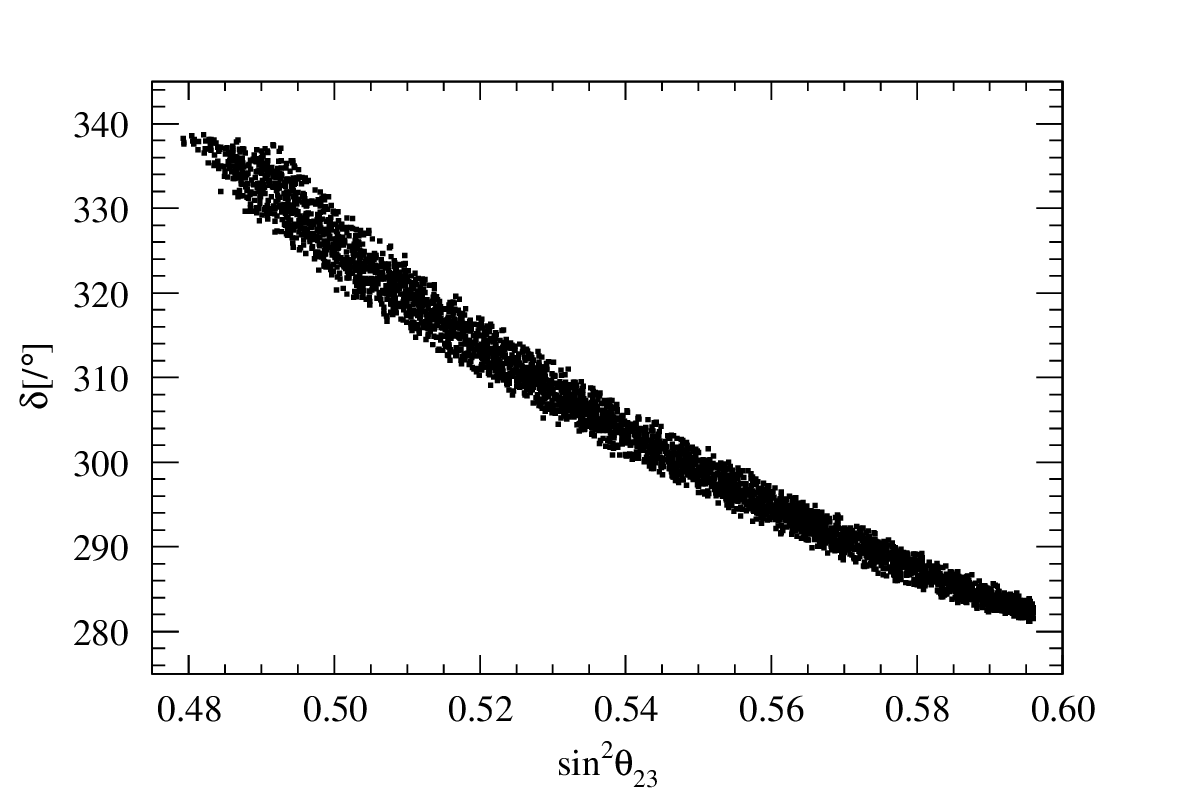}}
\hfill
\subfigure[Scenario II]{\label{fig1b}
\includegraphics[width=0.3\textwidth]{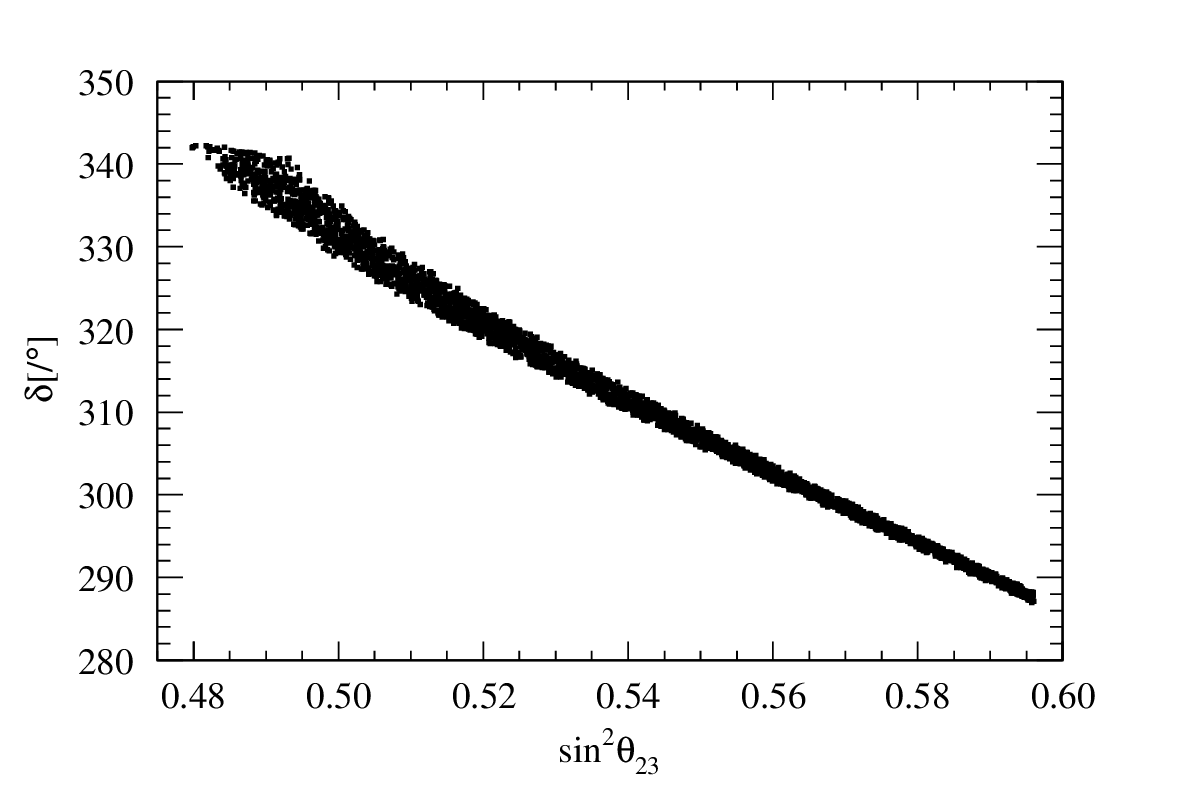}}
\hfill
\subfigure[Scenario III]{\label{fig1c}
\includegraphics[width=0.3\textwidth]{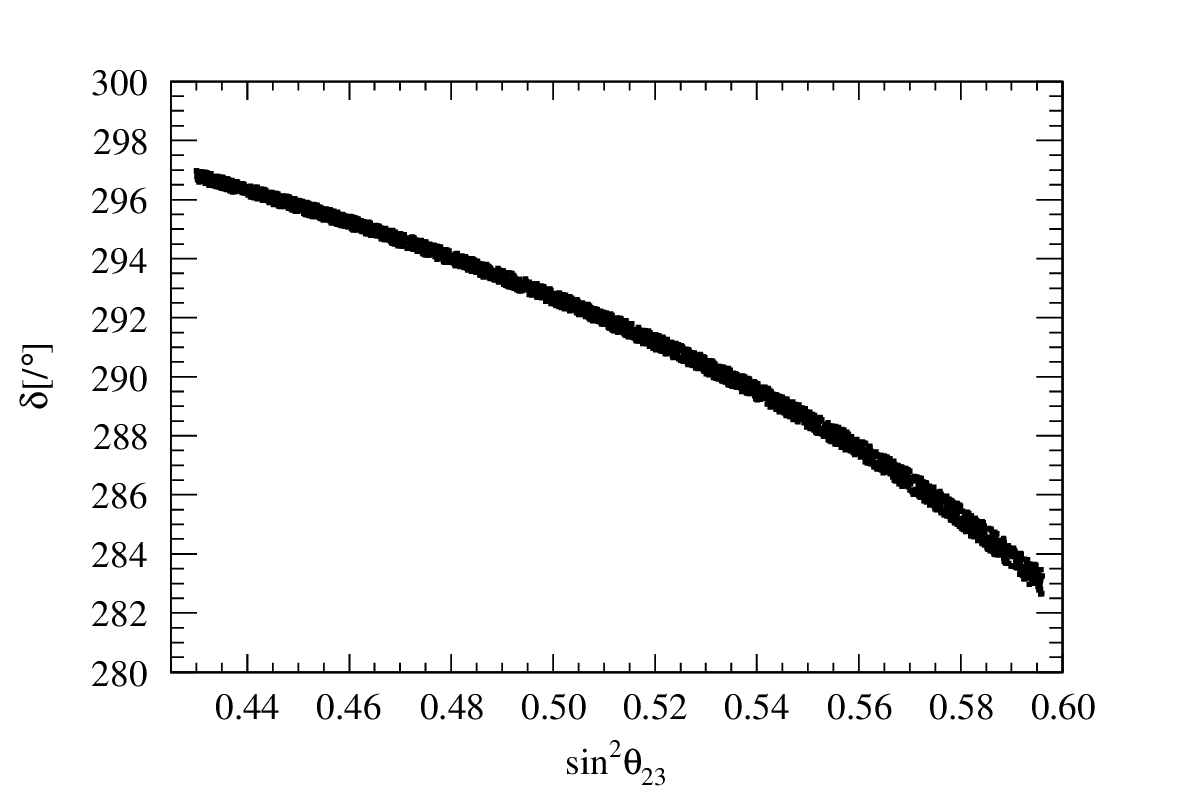}}
\caption{The predicted Dirac CP-violating phase $\delta$ as a function of $\sin^2\theta_{23}$ for (a) Scenario I, (b) Scenario II and (c) Scenario III.}\label{fig}
\end{figure}

\section{Conclusion}\label{sec:conclusion}

Flavor symmetry is one of the main ways to explain the lepton mass spectra and flavor mixing. In this work, we adopt a phenomenological approach and apply the \( S_{3L} \times S_{3R} \) flavor symmetry to the mass matrices of charged-leptons and neutrinos in a similar way.
To explain the realistic lepton mass hierarchy, the symmetry-breaking chain $S_{3L} \times S_{3R} \rightarrow S_{2L} \times S_{2R} \rightarrow \emptyset$ is introduced to both the charged-lepton and neutrino mass matrices.
For three residual subgroups \( S^{ij}_{2L} \times S^{ij}_{2R} \) (\( ij = 23, 13, 12 \)), we systematically analyze the various parameterizations of the lepton mass matrices, and derive the corresponding lepton mass spectra, neutrino mixing angles, and Dirac CP-violating phase.

To test the viability of the framework, a detailed numerical analysis is performed.
While the three symmetry-breaking chains impose different structural constraints on the mass matrices and lead to distinct lepton mass spectra and mixing matrices, all the three scenarios are found to be consistent well with the current experimental constraints. Specifically, there are nine parameters in the model, four parameters in the charged-lepton sector $(c_l, \delta_l, \epsilon_l,  \varepsilon_l)$ and the other five parameters in the neutrino sector $(c_\nu, r_\nu, \delta_\nu, \epsilon_\nu, \varepsilon_\nu )$. By inputting nine experimental observables, namely three charged-lepton masses (\( m_e, m_\mu, m_\tau \)), three neutrino masses (\( m_1, m_2, m_3 \)), and three mixing angles (\( \theta_{12}, \theta_{23}, \theta_{13} \)), we can fully determine all the model parameters, and make predictions for the Dirac CP-violating phase $\delta$.
For instance, with the latest best-fit values provided by NuFIT 6.0 (2024), the predicted Dirac CP-violating phase corresponding to three distinct symmetry-breaking chains is $\delta \approx 294.6^\circ,~302.3^\circ$ and $287.0^\circ$.
To better assess the viability and generality of the model, we further extend the ranges of the input observables to their full $3\sigma$ intervals and perform a comprehensive random scan over the model parameter space. The allowed range of $\delta$ is $281.2^\circ \rightarrow 338.7^\circ$, $287.0^\circ \rightarrow 342.2^\circ$ and $282.7^\circ \rightarrow 297.0^\circ$, respectively.

Through a comprehensive numerical analysis, it is further substantiated that the symmetry-breaking scheme presented in this paper is consistent well with the current experimental data, offering significant theoretical insights into the lepton masses, flavor mixing and CP violation within the context of flavor symmetries. Of course, the symmetry-breaking chain $S_{3L} \times S_{3R} \rightarrow S_{2L} \times S_{2R} \rightarrow \emptyset$ may also be extended to the quark sector, thereby providing valuable information for elucidating the mass spectra and flavor mixing of quarks.

\section*{Acknowledgements}
This work was supported in part by the National Natural Science Foundation of China (Grants No. 12105162, 12305106), the Natural Science Foundation of Shandong Province (Grants No. ZR2021QA058, ZR2021QA040) and the Youth Innovation Technology Project of Higher School in Shandong Province (2023KJ146).


\newpage

\begin{thebibliography}{0}


\bibitem{Glashow:1961tr}
S.~L.~Glashow,
Nucl. Phys. \textbf{22}, 579-588 (1961).

\bibitem{Weinberg:1967tq}
S.~Weinberg,
Phys. Rev. Lett. \textbf{19}, 1264-1266 (1967).

\bibitem{Salam:1968rm}
A.~Salam,
Conf. Proc. C \textbf{680519}, 367-377 (1968).

\bibitem{DiBari:2021fhs}
P.~Di Bari,
Prog. Part. Nucl. Phys. \textbf{122}, 103913 (2022).

\bibitem{Arbey:2021gdg}
A.~Arbey and F.~Mahmoudi,
Prog. Part. Nucl. Phys. \textbf{119}, 103865 (2021).

\bibitem{Crivellin:2023zui}
A.~Crivellin and B.~Mellado,
Nature Rev. Phys. \textbf{6}, no.5, 294-309 (2024).

\bibitem{Losada:2021bxx}
M.~Losada, Y.~Nir, G.~Perez and Y.~Shpilman,
JHEP \textbf{04}, 030 (2022).

\bibitem{SajjadAthar:2021prg}
M.~Sajjad Athar, S.~W.~Barwick, T.~Brunner, J.~Cao, M.~Danilov, K.~Inoue, T.~Kajita, M.~Kowalski, M.~Lindner and K.~R.~Long, \textit{et al.}
Prog. Part. Nucl. Phys. \textbf{124}, 103947 (2022).

\bibitem{Denton:2025jkt}
P.~B.~Denton,
arXiv:2501.08374 [hep-ph].

\bibitem{SNO:2025koj}
M.~Abreu \textit{et al.} [SNO+],
Phys. Rev. Lett. \textbf{135}, no.12, 121801 (2025).

\bibitem{Super-Kamiokande:2023jbt}
K.~Abe \textit{et al.} [Super-Kamiokande],
Phys. Rev. D \textbf{109}, no.9, 092001 (2024).

\bibitem{T2K:2023mcm}
K.~Abe \textit{et al.} [T2K],
Phys. Rev. D \textbf{108}, no.7, 072011 (2023).

\bibitem{NOvA:2025tmb}
S.~Abubakar \textit{et al.} [NOvA],
arXiv:2509.04361 [hep-ex].

\bibitem{Esteban:2024eli}
I.~Esteban, M.~C.~Gonzalez-Garcia, M.~Maltoni, I.~Martinez-Soler, J.~P.~Pinheiro and T.~Schwetz,
JHEP \textbf{12}, 216 (2024).

\bibitem{Capozzi:2025wyn}
F.~Capozzi, W.~Giar{\`e}, E.~Lisi, A.~Marrone, A.~Melchiorri and A.~Palazzo,
Phys. Rev. D \textbf{111}, no.9, 093006 (2025).

\bibitem{deSalas:2020pgw}
P.~F.~de Salas, D.~V.~Forero, S.~Gariazzo, P.~Mart{\'\i}nez-Mirav{\'e}, O.~Mena, C.~A.~Ternes, M.~T{\'o}rtola and J.~W.~F.~Valle,
JHEP \textbf{02}, 071 (2021).

\bibitem{KATRIN:2024cdt}
M.~Aker \textit{et al.} [KATRIN],
Science \textbf{388}, no.6743, adq9592 (2025).

\bibitem{Minkowski:1977sc}
P.~Minkowski,
Phys. Lett. B \textbf{67}, 421-428 (1977).

\bibitem{Yanagida:1979as}
T.~Yanagida,
Conf. Proc. C \textbf{7902131}, 95-99 (1979).

\bibitem{Gell-Mann:1979vob}
M.~Gell-Mann, P.~Ramond and R.~Slansky,
Conf. Proc. C \textbf{790927}, 315-321 (1979).

\bibitem{Glashow:1979nm}
S.~L.~Glashow,
NATO Sci. Ser. B \textbf{61}, 687 (1980).

\bibitem{Mohapatra:1979ia}
R.~N.~Mohapatra and G.~Senjanovic,
Phys. Rev. Lett. \textbf{44}, 912 (1980).

\bibitem{Wyler:1982dd}
D.~Wyler and L.~Wolfenstein,
Nucl. Phys. B \textbf{218}, 205-214 (1983).

\bibitem{Mohapatra:1986bd}
R.~N.~Mohapatra and J.~W.~F.~Valle,
Phys. Rev. D \textbf{34}, 1642 (1986).

\bibitem{Branco:1978bz}
G.~C.~Branco and G.~Senjanovic,
Phys. Rev. D \textbf{18}, 1621 (1978).

\bibitem{Chang:1986bp}
D.~Chang and R.~N.~Mohapatra,
Phys. Rev. Lett. \textbf{58}, 1600 (1987).

\bibitem{Babu:1988yq}
K.~S.~Babu and X.~G.~He,
Mod. Phys. Lett. A \textbf{4}, 61 (1989).

\bibitem{Hung:1998tv}
P.~Q.~Hung,
Phys. Rev. D \textbf{59}, 113008 (1999).

\bibitem{Xing:2020ijf}
Z.~z.~Xing,
Phys. Rept. \textbf{854}, 1-147 (2020).

\bibitem{King:2013eh}
S.~F.~King and C.~Luhn,
Rept. Prog. Phys. \textbf{76}, 056201 (2013).

\bibitem{Harari:1978yi}
H.~Harari, H.~Haut and J.~Weyers,
Phys. Lett. B \textbf{78}, 459-461 (1978).

\bibitem{Froggatt:1978nt}
C.~D.~Froggatt and H.~B.~Nielsen,
Nucl. Phys. B \textbf{147}, 277-298 (1979).


\bibitem{Koide:1989ds}
Y.~Koide,
Z. Phys. C \textbf{45}, 39 (1989).

\bibitem{Tanimoto:1989qh}
M.~Tanimoto,
Phys. Rev. D \textbf{41}, 1586 (1990).

\bibitem{Kaus:1990ij}
P.~Kaus and S.~Meshkov,
Phys. Rev. D \textbf{42}, 1863-1867 (1990).

\bibitem{Branco:1990fj}
G.~C.~Branco, J.~I.~Silva-Marcos and M.~N.~Rebelo,
Phys. Lett. B \textbf{237}, 446-450 (1990).

\bibitem{Fritzsch:1989qm}
H.~Fritzsch and J.~Plankl,
Phys. Lett. B \textbf{237}, 451-456 (1990).

\bibitem{Fritzsch:1994yx}
H.~Fritzsch and D.~Holtmannspotter,
Phys. Lett. B \textbf{338}, 290-294 (1994).

\bibitem{Branco:1995pw}
G.~C.~Branco and J.~I.~Silva-Marcos,
Phys. Lett. B \textbf{359}, 166-174 (1995).

\bibitem{Fritzsch:1995dj}
H.~Fritzsch and Z.~Z.~Xing,
Phys. Lett. B \textbf{372}, 265-270 (1996).

\bibitem{Xing:1996hi}
Z~z.~Xing,
J. Phys. G \textbf{23}, 1563-1578 (1997).

\bibitem{Mondragon:1998gy}
A.~Mondragon and E.~Rodriguez-Jauregui,
Phys. Rev. D \textbf{59}, 093009 (1999).

\bibitem{Fritzsch:1998xs}
H.~Fritzsch and Z.~z.~Xing,
Phys. Lett. B \textbf{440}, 313-318 (1998).

\bibitem{Haba:2000rf}
N.~Haba, Y.~Matsui, N.~Okamura and T.~Suzuki,
Phys. Lett. B \textbf{489}, 184-193 (2000).

\bibitem{Branco:2001hn}
G.~C.~Branco and J.~I.~Silva-Marcos,
Phys. Lett. B \textbf{526}, 104-110 (2002).

\bibitem{Fujii:2002jw}
M.~Fujii, K.~Hamaguchi and T.~Yanagida,
Phys. Rev. D \textbf{65}, 115012 (2002).

\bibitem{Fritzsch:2004xc}
H.~Fritzsch and Z.~z.~Xing,
Phys. Lett. B \textbf{598}, 237-242 (2004).

\bibitem{Rodejohann:2004qh}
W.~Rodejohann and Z.~z.~Xing,
Phys. Lett. B \textbf{601}, 176-183 (2004).

\bibitem{Teshima:2005bk}
T.~Teshima,
Phys. Rev. D \textbf{73}, 045019 (2006).

\bibitem{Altarelli:2010gt}
G.~Altarelli and F.~Feruglio,
Rev. Mod. Phys. \textbf{82}, 2701-2729 (2010).

\bibitem{Ishimori:2010au}
H.~Ishimori, T.~Kobayashi, H.~Ohki, Y.~Shimizu, H.~Okada and M.~Tanimoto,
Prog. Theor. Phys. Suppl. \textbf{183}, 1-163 (2010).

\bibitem{Xing:2010iu}
Z.~z.~Xing, D.~Yang and S.~Zhou,
Phys. Lett. B \textbf{690}, 304-310 (2010).

\bibitem{Zhou:2011nu}
S.~Zhou,
Phys. Lett. B \textbf{704}, 291-295 (2011).

\bibitem{Dev:2012ns}
S.~Dev, R.~R.~Gautam and L.~Singh,
Phys. Lett. B \textbf{708}, 284-289 (2012).

\bibitem{GonzalezCanales:2012blg}
F.~Gonzalez Canales, A.~Mondragon and M.~Mondragon,
Fortsch. Phys. \textbf{61}, 546-570 (2013).

\bibitem{Jora:2012nw}
R.~Jora, J.~Schechter and M.~N.~Shahid,
Int. J. Mod. Phys. A \textbf{28}, 1350028 (2013).

\bibitem{Ishimori:2013woa}
H.~Ishimori, T.~Kobayashi, Y.~Shimizu, H.~Ohki, H.~Okada and M.~Tanimoto,
Fortsch. Phys. \textbf{61}, 441-465 (2013).

\bibitem{Yang:2016esx}
M.~J.~S.~Yang,
Phys. Lett. B \textbf{760}, 747-752 (2016).

\bibitem{Fritzsch:2017tyf}
H.~Fritzsch, Z.~z.~Xing and D.~Zhang,
Chin. Phys. C \textbf{41}, no.9, 093104 (2017).

\bibitem{Si:2017pdo}
Z.~g.~Si, X.~h.~Yang and S.~Zhou,
Chin. Phys. C \textbf{41}, no.11, 113105 (2017).

\bibitem{Pramanick:2019oxb}
S.~Pramanick,
Phys. Rev. D \textbf{100}, no.3, 035009 (2019).

\bibitem{Mishra:2019sye}
S.~Mishra and A.~Giri,
J. Phys. G \textbf{47}, no.5, 055008 (2020).

\bibitem{Mishra:2019keq}
S.~Mishra,
Eur. Phys. J. Plus \textbf{135}, no.6, 485 (2020).

\bibitem{Garcia-Aguilar:2021xgk}
J.~D.~Garc{\'\i}a-Aguilar and J.~C.~G{\'o}mez-Izquierdo,
Rev. Mex. Fis. \textbf{68}, no.4, 040801 (2022).

\bibitem{Babu:2023oih}
K.~S.~Babu, Y.~Wu and S.~Xu,
JHEP \textbf{12}, 166 (2024).

\bibitem{Li:2024pff}
C.~C.~Li, J.~N.~Lu and G.~J.~Ding,
JHEP \textbf{12}, 015 (2024).

\bibitem{Gomez-Izquierdo:2024apr}
J.~C.~G{\'o}mez-Izquierdo, C.~Espinoza, L.~E.~G.~Luna and M.~Mondrag{\'o}n,
Nucl. Phys. B \textbf{1018}, 117027 (2025).

\bibitem{Gresnigt:2024awl}
N.~Gresnigt and L.~Gourlay,
J. Phys. Conf. Ser. \textbf{2912}, no.1, 012019 (2024).

\bibitem{ParticleDataGroup:2024cfk}
S.~Navas \textit{et al.} [Particle Data Group],
Phys. Rev. D \textbf{110}, no.3, 030001 (2024).

\bibitem{Planck:2018vyg}
N.~Aghanim \textit{et al.} [Planck],
Astron. Astrophys. \textbf{641}, A6 (2020)
[erratum: Astron. Astrophys. \textbf{652}, C4 (2021)].

\bibitem{Huang:2020hdv}
G.~y.~Huang and S.~Zhou,
Phys. Rev. D \textbf{103}, no.1, 016010 (2021).

\bibitem{Jarlskog:1985ht}
C.~Jarlskog,
Phys. Rev. Lett. \textbf{55}, 1039 (1985).

\bibitem{Wu:1985ea}
D.~d.~Wu,
Phys. Rev. D \textbf{33}, 860 (1986).



\end{thebibliography}
\end{document}